\documentclass[12pt,preprint]{aastex} 
\input{epsf}
\newcommand{\mpc}{h^{-1}{\rm Mpc}}

\newcommand{\kms}{{\rm km} \, {\rm s}^{-1}}

\newcommand{\gsim}{\mathrel{\hbox{\rlap{\hbox{\lower4pt\hbox{$\sim$}}}\hbox{$>$}}}}
\newcommand{\lsim}{\mathrel{\hbox{\rlap{\hbox{\lower4pt\hbox{$\sim$}}}\hbox{$<$}}}}

\shortauthors{Rojas et al.} 
\shorttitle{Properties of Void Galaxies}

\begin{document}
\title{Photometric Properties of Void Galaxies in the Sloan Digital
Sky Survey} \author{Randall R. Rojas\altaffilmark{1}, Michael S. Vogeley\altaffilmark{1},
Fiona Hoyle\altaffilmark{1}, and Jon Brinkmann\altaffilmark{2}}  
\email{rrojas@mercury.physics.drexel.edu, vogeley@drexel.edu,
hoyle@venus.physics.drexel.edu }

\altaffiltext{1}{Department of Physics,
Drexel University, 3141 Chestnut Street, Philadelphia, PA 19104}
\altaffiltext{2}{Apache Point Observatory, P.O. Box 59, Sunspot, 
NM 88349-0059}
 
\begin{abstract}
Using a nearest neighbor analysis, we construct a sample of void
galaxies from the Sloan Digital Sky Survey (SDSS) and compare the
photometric properties of these galaxies to the population of non-void
(wall) galaxies. We trace the density field of galaxies using a
volume-limited sample with $z_{max}=0.089$. Galaxies from the
flux-limited SDSS with $z\leq z_{max}$ and fewer than three
volume-limited neighbors within 7$h^{-1}$Mpc are classified as void
galaxies. This criterion implies a density contrast $\delta \rho/ \rho
< -0.6$ around void galaxies. From 155,000 galaxies, we obtain a
sub-sample of 13,742 galaxies with $z\leq z_{max}$, from which we
identify 1,010 galaxies as void galaxies. To identify an additional
194 faint void galaxies from the SDSS in the nearby universe, $r\lsim
72 \mpc$, we employ volume-limited samples extracted from the Updated
Zwicky Catalog and the Southern Sky Redshift Survey with
$z_{max}=0.025$ to trace the galaxy distribution. Our void galaxies
span a range of absolute magnitude from $M_r=-13.5$ to
$M_r=-22.5$. Using SDSS photometry, we compare the colors,
concentration indices, and Sersic indices of the void and wall
samples.  Void galaxies are significantly bluer than galaxies lying at
higher density. The population of void galaxies with $M_r \lsim M^*
+1$ and brighter is on average bluer and more concentrated (later
type) than galaxies outside of voids. The latter behavior is only
partly explained by the paucity of luminous red galaxies in
voids. These results generally agree with the predictions of
semi-analytic models for galaxy formation in cold dark matter models,
which indicate that void galaxies should be relatively bluer, more
disklike, and have higher specific star formation rates.
\end{abstract}


\section{Introduction}

\label{sec:intro}

Since the discovery of the void in Bo\"{o}tes (Kirshner et al. 1981),
with a diameter of $50\mpc$, and subsequent discoveries of voids in
larger redshift surveys (Geller \& Huchra 1989; Pellegrini, da Costa
\& de Carvalho 1989; da Costa et al. 1994; Shectman et al. 1996;
El-Ad, Piran \& da Costa 1996, 1997; M\"{u}ller et al. 2000; Plionis \&
Basilakos 2002; Hoyle \& Vogeley 2002), these structures have posed an
observational and theoretical challenge. Because the characteristic
scale of large voids was comparable to the depth of early redshift
surveys, few independent structures were detected, making statistical
analysis of their properties difficult. Likewise, the limitations of
computing technology constrained early cosmological simulations to
include only a few voids per simulation. 

Whether voids are empty or not has been the question of recent debate.
Peebles (2001) pointed out the apparent discrepancy between Cold
Dark Matter Models (CDM) and observations. CDM models predict
mass and hence, maybe galaxies, inside the voids, (Dekel \& Silk 1986;
Hoffman, Silk \& Wyse 1992). However, pointed observations toward void
regions failed to detect a significant population of faint galaxies
(Kuhn, Hopp \& Els\"{a}sser 1997; Popescu, Hopp \& Els\"{a}sser 1997;
McLin et al. 2002).  Surveys of dwarf galaxies indicate that they
trace the same overall structures as larger galaxies (Bingelli 1989).
Thuan, Gott \& Schneider (1987), Babul \& Postman (1990) and Mo,
McGaugh \& Bothun (1994) showed that galaxies had common voids
regardless of Hubble type.

Grogin \& Geller (1999, 2000) identified a sample of 149 galaxies that
lie in voids traced by the Center for Astrophysics Survey. The void
galaxies were found in the Century and 15R redshift samples.  Grogin
\& Geller showed that the void galaxies tended to be bluer and that a
significant fraction of them were of late type. Their sample of 149
void galaxies covered a narrow range of absolute magnitude ($-20\lsim
B\lsim-17$) of which 49 have a low density contrast of $\delta
\rho/\rho\leq-0.5$. Here we present a sample of $\sim10^{3}$ void
galaxies found in regions of density contrast
$\delta\rho/\rho\leq-0.6$. This sample is large enough to allow
comparison of void and wall galaxies with the same color, surface
brightness profile and luminosity to statistically quantify their
differences. The range of absolute magnitude (SDSS $r$-band) in our
sample ($-22\lsim M_{r}\lsim13$) is large enough to include faint dwarfs to 
giants.

In this paper, we introduce a new sample of void galaxies from the Sloan
Digital Sky Survey (SDSS). The large sky coverage and depth of the
SDSS provides us with the opportunity to identify for the first time
more than 10$^3$ void galaxies with $\delta \rho/\rho\leq-0.6$. In
Section \ref{sec:surv} we discuss the galaxy redshift samples that we
use for this analysis. In Section \ref{sec:fvgs} we describe our
method for finding void galaxies. In Section \ref{sec:props} we
present the results found from the comparison of the photometric
properties of void and wall galaxies and in Section \ref{sec:discus}
we interpret these results by comparing them to predictions from
semi-analytic modeling of structure formation and properties of
different galaxy types. Finally, in Section \ref{sec:conc} we present
our conclusions.

\section{The Redshift Surveys}
\label{sec:surv}

The search for void galaxies requires a large 3-dimensional map of the
galaxy density field. We extract a volume-limited sample from the SDSS
data to map the galaxy density field and look for void galaxies in the
full magnitude-limited sample. As the SDSS currently has a slice-like
geometry, with each slice only $\sim 6^{\circ}$ thick, large voids of
radius $\sim 10\mpc$ ($h\equiv \ H_{o}/100 {\rm km}\,\, {\rm s}^{-1}
{\rm Mpc}^{-1}$) can only be detected at comoving distances of $r\gsim
10^2\mpc$ using the SDSS data alone. Therefore, to trace the local
voids, we also extract a volume-limited sample from the combined
Updated Zwicky Catalog (UZC; Falco et al. 1999) and Southern Sky
Redshift Survey (SSRS2; da Costa et al. 1998). It should be noted that
nearby void galaxies are not selected from the UZC and SSRS2 surveys.
These surveys are only used to define the density field around SDSS
galaxies that lie at distances $r \leq72\mpc$.

To recap, we have two volume-limited samples, one from the SDSS and
one from the combined UZC+SSRS2. These samples are used to define the
galaxy density field only. Void galaxies are found from the
magnitude-limited SDSS sample. We define the Distant sample to be the
SDSS magnitude-limited sample truncated at $100\leq r
\leq260\mpc$. The Nearby sample is the SDSS magnitude-limited sample
truncated at $r=72\mpc$. Both magnitude-limited samples (Nearby and
Distant) are constructed using the SDSS $r$-band. In this section we
describe each of the surveys and samples in detail.

\subsection{The SDSS}
\label{sec:SDSS}

The SDSS is a wide-field photometric and spectroscopic survey. The
completed survey will cover approximately $10^{4}$ square degrees.
CCD imaging of 10$^8$ galaxies in five colors and follow-up
spectroscopy of 10$^6$ galaxies with $r<17.77$ will be
obtained.  York et al. (2000) provides an overview of the SDSS and
Stoughton et al. (2002) describes the early data release (EDR) and
details about the photometric and spectroscopic measurements made from
the data. Abazajian et al. (2003) describes the First Data Release
(DR1) of the SDSS. Technical articles providing details of the SDSS
include descriptions of the photometric camera (Gunn 1998),
photometric analysis (Lupton et al. 2002), the photometric system
(Fukugita et al. 1996; Smith et al. 2002), the photometric monitor
(Hogg et al. 2001), astrometric calibration (Pier et al. 2002),
selection of the galaxy spectroscopic samples (Strauss et al. 2002;
Eisenstein et al. 2001), and spectroscopic tiling (Blanton et
al. 2001).  A thorough analysis of possible systematic uncertainties
in the galaxy samples is described in Scranton et al. (2002).

We examine a sample of 155,126 SDSS galaxies (Blanton et al. 2002;
{\tt sample10}) that have both completed imaging and spectroscopy. The
area observed by {\tt sample10} is approximately 1.5 times that of the
DR1 (Abazajian et al. 2003). To a good approximation, the sample we
analyze consists of roughly three regions covering a total angular
area of 1,986 deg$^2$. Due to the complicated geometry of the SDSS sky
coverage, the survey regions are best described in the SDSS coordinate
system (see Stoughton et al. 2002). Where possible in this section we
describe approximate limits in the more familiar equatorial
coordinates. The first region is an equatorial stripe in the North
Galactic Cap (NGC). This stripe has a maximum extent of $7.5^{\circ}$ in
the declination direction over the R.A.  range $21^{\rm h}30^{\rm m}
\lsim\alpha\lsim4^{\rm h}10^{\rm m}$ and maximum length of
$120^{\circ}$ over the R.A. range $8^{\rm h}\lsim\alpha\lsim16^{\rm
h}$. The second region is in the South Galactic Cap (SGC). There are
three stripes, the boundaries of which are defined in the SDSS
coordinate system. Each stripe is $2.5^{\circ}$ wide in SDSS survey
coordinates. One stripe is centered at $\delta = 0^{\circ}$ and covers
the R.A.  range $20^{\rm h}40^{\rm m}\lsim\alpha\lsim2^{\rm h}20^{\rm
m}$. The other two stripes are above and below the equator and cover
similar R.A. ranges. In survey coordinates these two stripes cover the range
$-28^{\circ}\lsim\lambda\lsim41^{\circ}$,
$130^{\circ}\lsim\eta\lsim135^{\circ}$ and
$-57^{\circ}\lsim\lambda\lsim58^{\circ}$,
$155^{\circ}\lsim\eta\lsim160^{\circ}$.  The third large region is in
the North Galactic Cap. In SDSS survey coordinates it covers the range
$-48^{\circ}\lsim\lambda\lsim50^{\circ}$,
$75^{\circ}\lsim\eta\lsim85^{\circ}$.  There are additional smaller
stripes at $-37^{\circ}\lsim\lambda\lsim-22^{\circ}$,
$60^{\circ}\lsim\eta\lsim70^{\circ}$ and
$2^{\circ}\lsim\lambda\lsim60^{\circ}$,
$90^{\circ}\lsim\eta\lsim100^{\circ}$ (the boundary is an
approximation because of the tiling geometry).

We correct the velocities of galaxies to the Local Group frame
according to
\begin{equation}
\Delta v=V_{\rm apex}[\sin(b)\sin(b_{\rm apex})+\cos(b)\cos(b_{\rm
  apex})\cos(l-l_{\rm apex})]
\end{equation}
where $l_{\rm apex}=93^\circ$, $b_{\rm apex}=-4^\circ$, and $V_{\rm
apex}=316 \, \kms$ (Karachentsev \& Makarov 1996). The magnitudes of
the galaxies are $K$-corrected as described in Blanton et al. (2003)
and corrections for Galactic extinction are made using the Schlegel,
Finkbeiner, \& Davis (1998) dust maps. Finally, to convert redshifts
into comoving distances we adopt an $(\Omega_{\rm
m},\Omega_{\Lambda})=(0.3,0.7)$ cosmology.

The decrease of observed galaxy density with distance in an apparent
magnitude-limited galaxy sample might cause us to erroneously detect
more voids at large distances. Therefore, we use a volume-limited
sub-sample of the SDSS to define the density field of galaxies. This
sample consists of galaxies with redshifts less than the redshift
limit, $z_{max}$, and SDSS $r$-band absolute magnitudes brighter than
$M_{\rm crit}$, where
\begin{equation}
M_{\rm crit} = r_{\rm lim} - 25 - 5 {\rm log_{10}}[d_l(\rm z_{max})]
\end{equation}
$r_{\rm lim}$\footnote{We use $r_{\rm lim}=17.5$ instead of $r_{\rm
lim}=17.77$, in the construction of volume-limited catalog to ensure
we have a uniform limit across all the data since earlier stripes were
only observed to $r_{\rm lim}=17.5$.}, is the magnitude limit of the
survey and $d_l$, is the luminosity distance in units of $\mpc$ at
$\rm z_{max}$.  We form a volume-limited sample of the SDSS with $\rm
z_{max}=0.089$, with corresponding absolute-magnitude limit $M_{\rm
crit}=-19.87$ (in the SDSS $r$-band).  The redshift limit $\rm
z_{max}=0.089$ allows us to construct the largest possible
volume-limited sample from the current SDSS sample. This
volume-limited sample contains 22,866 galaxies where the mean
separation between these galaxies is $\sim5.3 \mpc$. For a
$(\Omega_{\rm m}=0.3, \Omega_{\Lambda}=0.7)$ cosmology, the redshift
limit of $z_{\rm max}$=0.089, corresponds to a comoving distance of
$260\mpc$. The lower bound of $100\mpc$ on the comoving distance is
necessary due to the slice-like geometry of the early SDSS
slices. Recall that voids of diameter $\gsim10\mpc$ can only be found at
$r\gsim100\mpc$ as discussed in Section 2.

\subsection{The Updated Zwicky Catalog}

The Updated Zwicky Catalog (Falco et al. 1999) includes a re-analysis
of data taken from the Zwicky Catalog and Center for Astrophysics
surveys (Zwicky et al. 1961-1968; Geller \& Huchra 1989; Huchra et
al. 1990; Huchra, Geller, \& Corwin 1995; Huchra, Vogeley, \& Geller
1999) together with new spectroscopic redshifts for some galaxies and
coordinates from the digitized POSS-II plates. Improvements over the
previous catalogs include estimates of the accuracy of the CfA
redshifts and uniformly accurate coordinates at the $< 2^{\prime
\prime}$ level.

The UZC contains 19,369 galaxies. Of the objects with limiting
apparent magnitude $m_{\rm Zw} \leq 15.5$, 96\% have measured
redshifts, giving a total number of 18,633 objects. The catalog covers
two main survey regions: $8^{h}<\alpha_{1950}<17^{h},
8.5^{\circ}<\delta_{1950}<44.5^{\circ}$ in the North Galactic Cap and
$20^{h}<\alpha_{1950}<4^{h}, -2.5^{\circ}<\delta_{1950}<48^{\circ}$ in
the South Galactic Cap. We correct the velocities of the galaxies with
respect to the Local Group as discussed in Section 2.1. The magnitudes
of the galaxies are corrected for Galactic extinction using the
Schlegel, Finkbeiner \& Davis (1998) dust maps and the magnitudes are
$K$-corrected assuming $K=3$, which is appropriate for the {\it B}
filter and the median galaxy morphological type Sab (Park et al. 1994;
Pence 1976; Efstathiou, Ellis \& Peterson 1988).

We construct a volume-limited UZC sample with ${\rm z_{max}}=0.025$
since this is the redshift at which the largest volume-limited sample
can be obtained. This volume-limited sample contains 4924 galaxies,
has a comoving depth of $\sim 76 \mpc$ and absolute-magnitude limit
of $M_{lim}\leq-18.96$ ($B_{\rm Zw}$). To compare this limit to that
of the SDSS, we translate a B-band magnitude into an approximate
$r$-band magnitude of $M_r=-20.06$ using $g-r=0.66$ and $g-B=-0.45$ from
Fukugita et al. (1995). The absolute magnitude limit of the UZC sample
is therefore, slightly brighter than the SDSS limit. To ensure that
this sample and the SSRS2 described below are equally deep, we cut
back this sample to $72\mpc$.

\subsection{The SSRS2}

The SSRS2 galaxy sample (da Costa et al. 1998) was selected from the
list of nonstellar objects in the Hubble Space Telescope Guide Star
Catalog (GSC). The SSRS2 contains 3489 galaxies in the SGC over the
angular region: $\delta_{1950}\leq-2.5^{\circ}$ and
$b\leq-40^{\circ}$, covering a total of 1.13 sr with $m_{\rm
SSRS2}\leq 15.5$, where the zero-point offset from the Zwicky
magnitude system used in the UZC is approximately $m_{\rm
SSRS2}-m_{\rm Zwicky}\sim 0.10$ mag (Alonso et al. 1994).

We construct a volume-limited sample with the same redshift limit as
for the UZC, ${\rm z_{max}}=0.025$ (same reason as discussed in
Section 2.2) and (after adjustment of the zeropoint),
$M\leq-18.96$. For our chosen cosmology, the depth of the sample is
$\sim 73 \mpc$ which we also cut back to $72\mpc$ as discussed in the
case for the UZC sample. Therefore, both SSRS2 and UZC volume-limited
samples have the same comoving depth.

As above (Section 2.1), we correct galaxy
velocities to the Local Group frame, apply the Galactic dust
corrections based on the Schlegel, Finkbeiner, \& Davis (1998) dust
maps, and assume $K=3$ to $K$-correct the observed magnitudes. This
volume-limited sample includes 725 galaxies.

The SSRS2 provides angular coverage in the South Galactic Cap. The
combined UZC+SSRS2 sample contains 5649 galaxies and sky coverage of
$\sim4.25\, \rm {sr}$.

\subsection{Summary of Surveys}

\begin{figure} 
\begin{centering}
\begin{tabular}{cc}
{\epsfxsize=10truecm \epsfysize=10truecm \epsfbox[0 170 650 600]{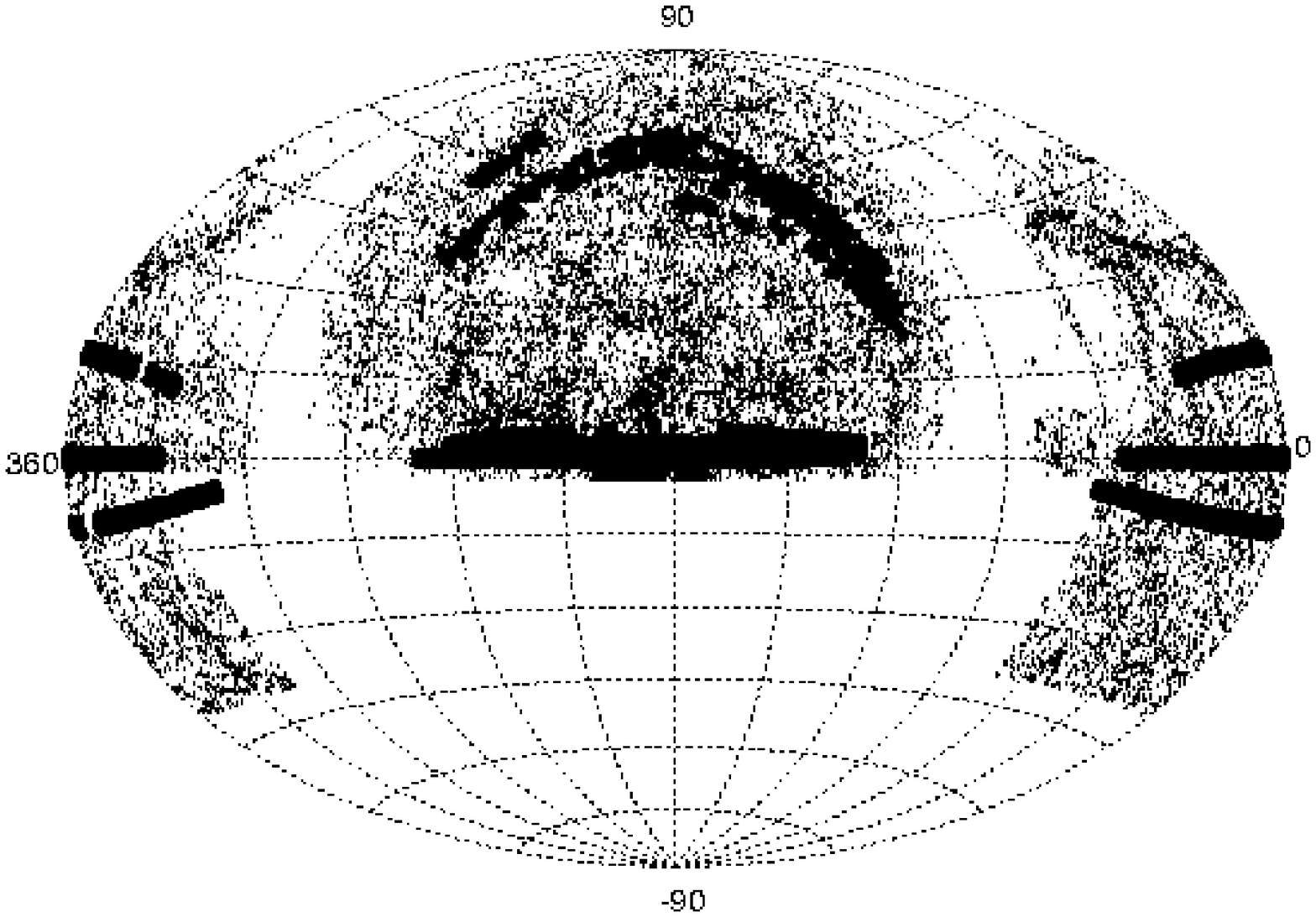}}&
{\epsfxsize=7truecm \epsfysize=7truecm \epsfbox[35 170 550 675 ]{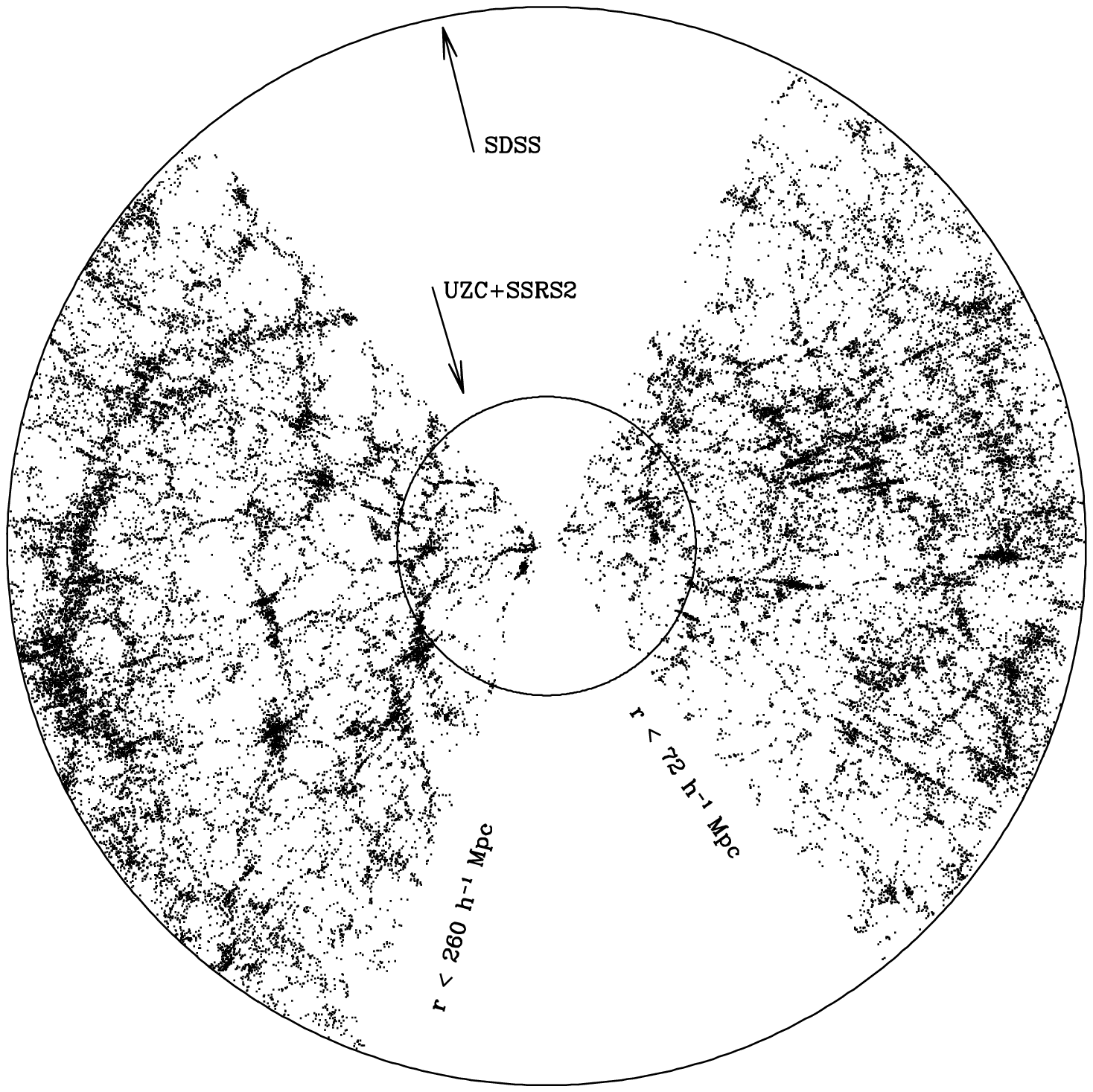}}
\end{tabular}
\caption{The left plot shows an Aitoff projection in celestial
coordinates of the current SDSS galaxy redshift survey data (black
points) and the combined Updated Zwicky Catalog and Southern Sky
Redshift Survey (gray dots). Approximate coordinates of each region
are given in the text in Section \ref{sec:surv}. The right plot shows
a cone diagram of the flux-limited SDSS data with $|\delta| \lsim
15^{\circ}$. The inner circle is drawn at $72\mpc$, which is the depth
of the combined UZC and SSRS2 volume-limited sample. The outer circle
is drawn at $260\mpc$, which is the depth of the SDSS volume-limited
sample. In the region $100 < r < 260 \mpc$ we use the SDSS data to
trace the distribution of the voids. However, nearby the SDSS
currently is limited in volume as only narrow strips have been
observed. Therefore, nearby we use the UZC+SSRS2 to trace the voids out to $72 \mpc$.}
\label{fig:surveys}
\end{centering}
\end{figure}

The left hand plot of Figure \ref{fig:surveys}, shows an Aitoff
projection of the three surveys. The black points show the SDSS
galaxies and the gray dots show the UZC+SSRS2 galaxies. This figure
demonstrates that in terms of area, the SDSS is almost totally
embedded in the UZC+SSRS2 data apart from along the bottom edge of the
northern equatorial slice and a small part of the southern most
slice. Therefore, the combined UZC+SSRS2 survey is useful for defining
the large-scale galaxy density field around the SDSS sample out to a
distance of approximately $72\mpc$.

The right-hand plot in Figure 1 shows a cone diagram of the SDSS data
with $|\delta| \lsim 15^{\circ}$.  The inner circle is drawn at
$72\mpc$, which is the comoving depth of the combined UZC and SSRS2
volume-limited sample. The outer circle is drawn at $260\mpc$, which
is the comoving depth of the SDSS volume-limited sample. Beyond
$72\mpc$, the selection function (number of observed galaxy density with
distance) of these shallower surveys drops and the thickness (in the
declination direction) of the SDSS itself is adequate to define the
density field around the SDSS galaxies.

\section{Identifying Void Galaxies Using Nearest Neighbor Statistics}
\label{sec:fvgs}

We search for void galaxies in the SDSS using the nearest neighbor
statistic. The two volume limited samples (SDSS and UZC+SSRS2) are
used to trace the voids. Any magnitude-limited galaxy that lies away from
the boundary of the volume-limited sample and has less than 3
volume-limited sample neighbors in a sphere of $7\mpc$ is considered a
void galaxy. We expand on each of these steps below.

\subsection{Proximity to Survey Boundary}


Galaxies in the magnitude-limited SDSS samples that lie near the
boundaries of the volume-limited samples have systematically larger
distances to their third nearest neighbors than galaxies that lie deep
in the volume-limited samples. This is because potentially closer
neighbors have not been observed/included in the sample. These
galaxies have a higher probability of being selected as void galaxies
than the galaxies inside the survey. We correct for this bias in the
following way: We generate a random catalog with the same angular and
distance limits as the corresponding volume-limited sample
(SDSS and UZC+SSRS2) but with no clustering. We count how many random
points lie around each of the magnitude-limited SDSS galaxies. If the
density around a galaxy is less than a certain value, we reject it
from the SDSS samples. This is explained further below.

We count how many random points ($N$) lie in a sphere of size,
$r=3.5\mpc$ around each galaxy and compute the number density,
$\rho=N/V$, where, $V=(\frac{4 \pi}{3} r^{3})$. Since we know the
number of random points, the solid angle and depth of the SDSS and
UZC+SSRS2 surveys, we can compute the corresponding average density of
random points, $\rho_{random}=N/\frac{\Omega}{3} r^{3}$. Galaxies with
values of $\rho < \rho_{\rm random}$, are rejected as it is their
proximity to the sample's boundaries which causes a low value of
$\rho$. 

We apply the above procedure twice, once when we compare the distant SDSS
magnitude-limited sample with the SDSS random catalog and again when
we compare the nearby SDSS magnitude-limited sample with the UZC+SSRS2
random catalog. The distant SDSS sample is reduced from 65,186
galaxies to 13,742 galaxies, the nearby SDSS sample is reduced
from 3784 galaxies to 2,450 galaxies. The nearby SDSS sample is cut
less drastically as the UZC+SSRS2 sample covers a greater area.

Because the SDSS is not finished, the angular selection function is
complicated (see Figure \ref{fig:surveys}). An algorithm to quantify
the fraction of galaxies that have been observed in any given region,
i.e.  the completeness, has been developed and is described in
Tegmark, Hamilton, \& Xu (2002). The completeness for any given
($\alpha, \delta$) coordinate is returned, allowing a random catalog
with the same angular selection function to be created. For the SDSS,
the completeness within the regions that have been observed is
typically $> 90\%$. The angular selection function for the combined
UZC+SSRS2 sample is easier as the surveys are finished and the
completeness for the UZC is $\sim96\%$. 

\subsection{Nearest Neighbor}
\label{sec:nn}

We classify galaxies that have a large distance to their {\it nth}
nearest neighbor as void galaxies. We follow the work of El-Ad \&
Piran (1997) and Hoyle \& Vogeley (2002) and use $n=3$ rather than
$n=1$ in the nearest neighbor analysis. Because galaxies are
clustered, it is not unreasonable to expect that some galaxies in
large-scale voids might be found in binaries or triplets. If we used
$n=1$ then a pair of galaxies in an otherwise low density environment
would not be classified as void galaxies. Setting $n=3$ allows for a
couple of bright neighbors, but excludes galaxies in typical
groups. Note that we do not make any corrections for peculiar
velocities along the line of sight. Therefore, we might underestimate
the density of systems with large velocity dispersions, which may
pollute the void galaxy population. This effect could lead us to
slightly underestimate the differences between the void and wall
populations.

To identify void galaxies in the SDSS, we compute the distance from
each galaxy in the apparent magnitude limited sample to the third
nearest neighbor in a volume-limited sample. In other words, the
volume-limited sample is used to define the galaxy density field that
traces voids and other structures.  We compute the average distance to
the third nearest neighbor, $\bar{d}_{sep}^{(3)}$, and the standard
deviation, $\sigma_{sep}^{(3)}$, of this distance. We fix the critical
distance $d_{\rm crit}$ to be $7\mpc$, which is approximately equal to
$\bar{d}_{sep}^{(3)}+1.5 \sigma_{sep}^{(3)}$ found from the two
samples, (the actual values are given in Sections \ref{sec:DVG} and
\ref{sec:NVG}). This threshold is consistent with the criterion for
defining wall and void galaxies in {\tt VOIDFINDER} (Hoyle \& Vogeley
2002). Galaxies in the apparent magnitude limited sample whose third
nearest neighbor lies further than $\geq d_{\rm crit}=7\mpc$ are
classified as void galaxies.  We thereby divide the apparent magnitude
limited SDSS sample into two mutually-exclusive sub-samples, which we
hereafter refer to as void and wall galaxies.

\begin{figure}
\begin{centering}
\begin{tabular}{lc}
{\epsfxsize=7truecm \epsfysize=7truecm \epsfbox[0 170 650 600 ]{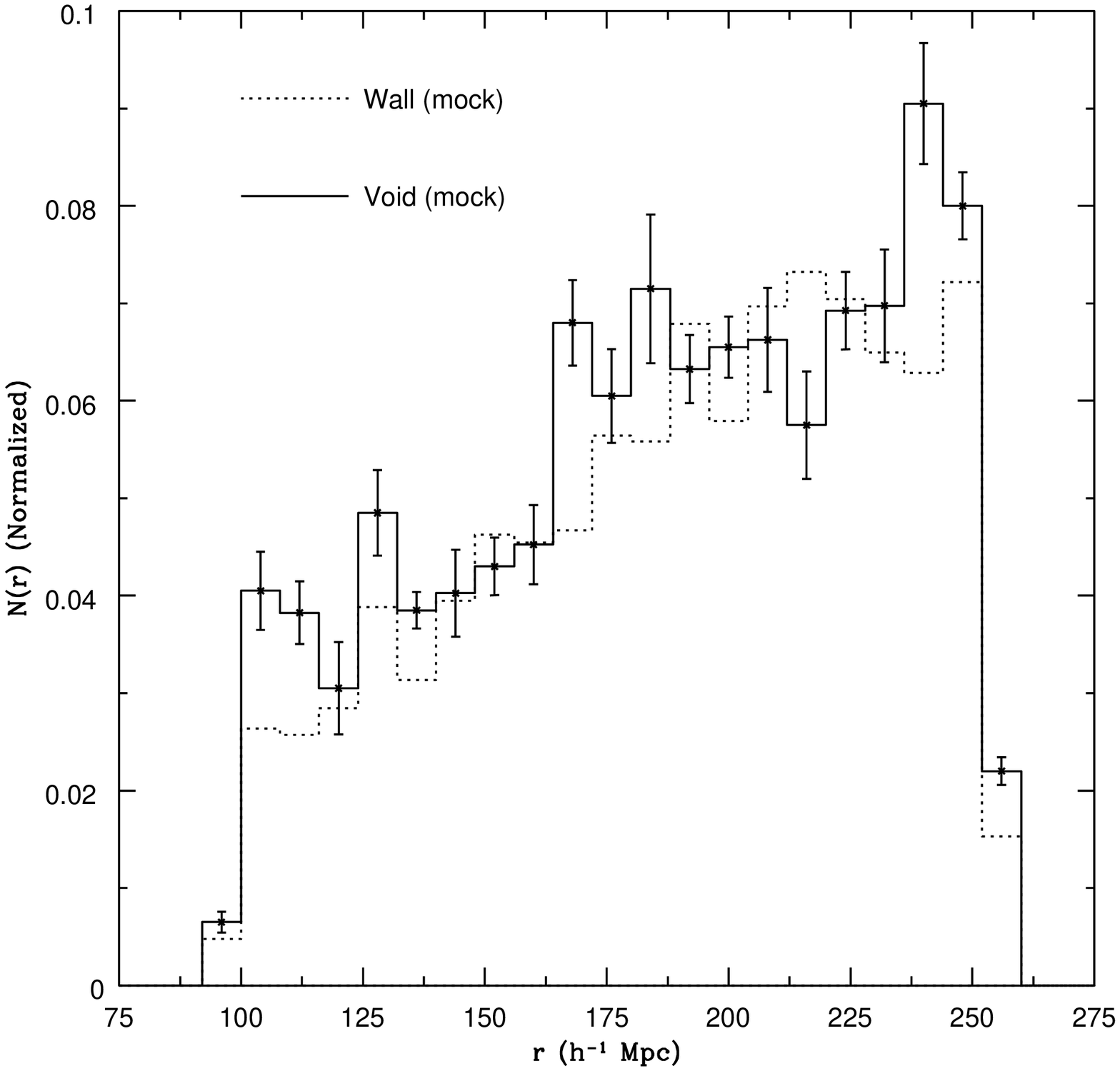}}&
{\epsfxsize=7truecm \epsfysize=7truecm \epsfbox[0 170 650 600]{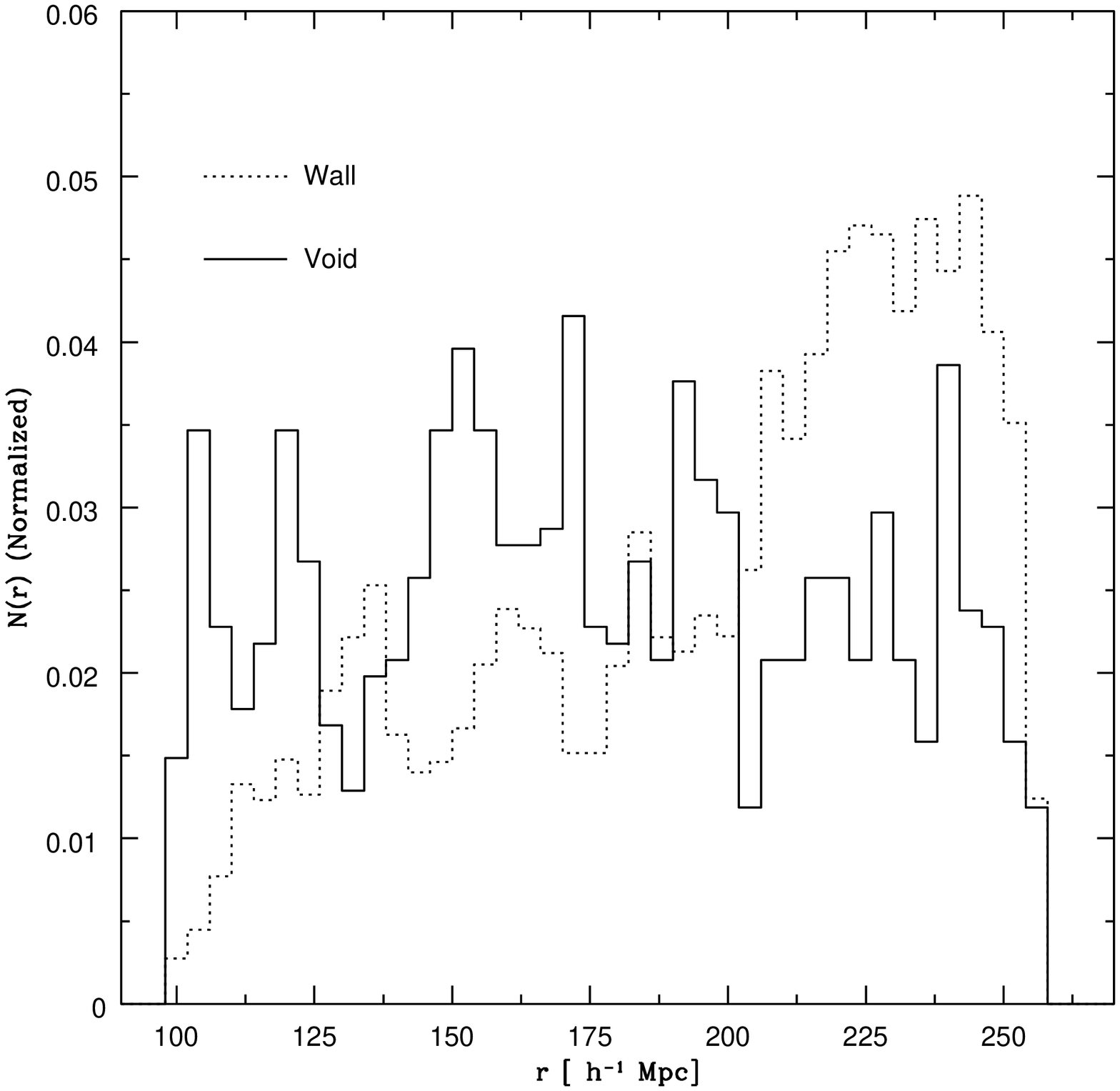}}
\end{tabular}
\caption{N(r) distribution of void and wall galaxies from mock
catalogs (left) and data (right). The left plot shows the N(r)
distribution as a function of comoving distance for the void (solid
line) and wall (dotted line) mock samples. The mock samples are
averaged over 10 independent realizations of region 2 (equatorial
stripe in the North Galactic Cap, see details in Section 2.1) from the
Virgo Consortium's Hubble Volume z=0 $\Lambda$CDM simulation (Frenk et
al. 2000).  The error bars are the $1 \sigma$ errors on the mock void
galaxy samples. The right plot shows the same distribution for the
distant void (solid line) and wall (dotted line) galaxy samples.}
\label{fig:nrdata}
\end{centering}
\end{figure}

Note that boundary of the void and wall samples is defined by throwing
away galaxies that lie within 3.5$h^{-1}$Mpc of the survey edge, where
as galaxies are classified as void galaxies if they have less than 3
neighbors in a sphere of 7$h^{-1}$Mpc. If we use 7h$^{-1}$Mpc to mark
the boundary then the volume available for finding void galaxies is
decreased, especially at the near edge of the distant sample. We
tolerate this inconsistancy inorder to have overlap between the near
and distant samples in terms of the magnitude ranges that each sample
probes. However, this means that near the edges there is a slightly
higher probability of a galaxy being flagged as a void galaxy than
deep in the survey. To test what effect this has, we construct 10 mock
volume- and flux-limited catalogs from the Virgo Consortium's Hubble
Volume z=0 $\Lambda$CDM simulation (Frenk et al. 2000; Evrard et
al. 2002) that have the same geometry as region 2. Following the
procedure used in the survey data, we throw away galaxies in the
flux-limited mock catalogs that lie within 3.5$h^{-1}$Mpc of the
region's edge and then find the mock void galaxies.  We then compute
the average N(r) distribution of the mock void and wall samples, where
N(r) is the normalized number of galaxies as a function of comoving
distance, and show these in the left hand plot of Figure
\ref{fig:nrdata}. It can be seen that within the errors, the two
distributions are similar. The exception is at the near edge where
more mock galaxies are classified as void galaxies, as expected. Out
to 125$h^{-1}$Mpc there are 50\% more void galaxies than wall
galaxies. This excess is only 4\% of the whole void sample because
most of the void galaxies are found at greater distances.  The N(r)
plot for the data (Figure \ref{fig:nrdata} right hand side) indicates
a somewhat later ratio of void/wall galaxies at the near edge of the
sample. From the test above, only 4\% of the void galaxies might be
erroneously flagged. The rest of the difference is due to large-scale
structure within the volume surveyed.

Thus we conclude that our procedure for identifying void galaxies and
 removing objects (both void and wall) near the survey boundaries does
 not produce any significant bias in the redshift distribution of void
 and wall galaxies. The difference is insufficient to generate the
 large observed differences between void and wall galaxies. In fact,
 dilution of the void galaxy sample can only decrease the apparent
 statistical significance of differences between the void and wall
 galaxy populations i.e. the true differences between void and wall
 galaxies may be more severe than we find. The converse is not
 possible; this dilution could not cause the population differences
 that we observe. In Section \ref{sec:discus} we discuss the impact of
 this dilution on our results. Also, tests with smaller "clean"
 samples show, as expected, a higher statistical significance and
 results from the nearby sample, which suffers less from this effect
 due to wider opening angle, and distant sample show the same trends,
 which is further evidence that the dilution of the distant sample is
 a minor effect.

To test that our procedure identifies genuine void galaxies, we
compute the mean, median and upper bound of the density contrast
($\delta \rho/\rho$) around void galaxies and compare these values to
the emptiness of voids as defined by {\tt VOIDFINDER}. The number of
galaxies in the SDSS volume-limited sample is 22,866 and the
respective volume, $V=\frac{\Omega}{3}(r^{3}-r_{o}^{3})=
\frac{0.6}{3}(260^3-100^3) = 3.32\times 10^6 h^{-3} {\rm Mpc}^3$,
therefore, the mean density is $\bar{\rho}=6.84\times 10^{-3} h^{3}
{\rm Mpc}^{-3}$. The void galaxies contain less than three neighbors
in a sphere of $7\mpc$, thus, the density around the void galaxies is
$ \rho_{v}\leq4/(\frac{4 \pi}{3}) 7^{3}= 2.78\times 10^{-3} h^{3} {\rm
Mpc}^{-3}$. Therefore, the density contrast around void galaxies in
the distant sample is $(\rho_v - \bar{\rho})/\bar{\rho} \leq
-0.6$. This number is very similar for the nearby sample. It is an
upper bound, as the median third nearest neighbor distance to the void
galaxies is closer to $8\mpc$, giving values for the density contrast
closer to $\delta \rho / \rho = -0.8$. This value is low, although not
as low as that found by {\tt VOIDFINDER} for the density contrast of
the voids. Since we are centered on a galaxy and galaxies are
clustered, we expect the density around void galaxies ($\rho_{vg}$) to
be higher than the mean density of a void
($\bar{\rho}_{void}$). Recall that the mean density of a void is about
$0.1\times$ mean density of the universe ($\bar{\rho}_{universe}$) and
since the correlation length ($\xi$) on spheres of $8\mpc$ is $\sim1$
($\sigma_{8}\sim1$), then:
$\rho_{vg}(r\leq7\mpc)=\bar{\rho}_{void}\times (1+\xi(r=7\mpc))\sim
2\times \bar{\rho}_{void}$. In addition, void galaxies are typically
found near the the edge of the void where the density is higher.

It is important to keep in mind that since most of the void galaxies
will lie near the edges of voids, the typical density contrast around
void galaxies is less extreme than the density contrast of the whole
void region (see Figure 11 in Benson et al. 2003). The average number
of volume-limited galaxies in a sphere of $7\mpc$ around a wall galaxy
is 25 compared to 2 around a void galaxy, demonstrating that void
galaxies really are in highly underdense regions.

\subsection{Distant Void Galaxies}
\label{sec:DVG}

For SDSS galaxies that lie in the distant SDSS sample, we use the SDSS
volume-limited sample to define the galaxy density field.  Using the
third nearest neighbor ($n=3$), we obtain
$(\bar{d}_{sep}^{(3)},\sigma_{sep}^{(3)})=(3.6, 2.10) \mpc$, from
which we obtain $d_{\rm crit}=6.75\mpc$, which we round up to
$7\mpc$. From the distant SDSS sample of 13,742 galaxies we find 1010
void galaxies. This sample of void galaxies will be referred to as VGD
(as in Void Galaxy Distant). The sample of 12,732 non-void galaxies we
label WGD (as in Wall Galaxy Distant). The fraction of void galaxies
in the distant sample is $\sim8\%$. This is only slightly higher than
the fraction of void galaxies found by {\tt VOIDFINDER} (Hoyle \&
Vogeley 2002) and by El-Ad \& Piran (1997).

\begin{figure}
\begin{centering}
\begin{tabular}{lc}
{\epsfxsize=10truecm \epsfysize=10truecm \epsfbox[35 170 550 675 ]{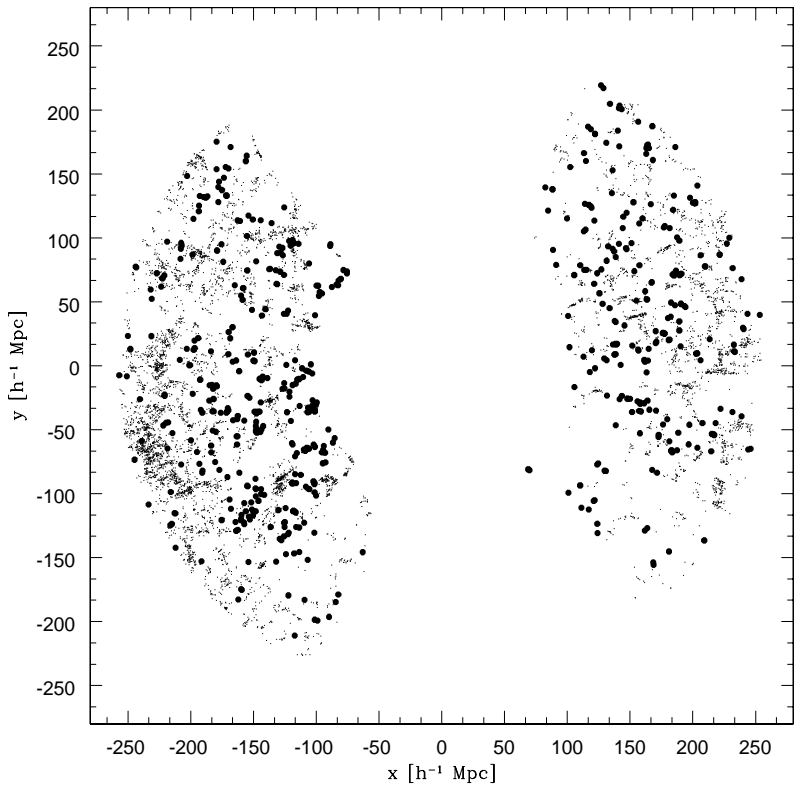}} 
\end{tabular}
\caption{Redshift space distribution of void and wall
galaxies. The gray dots show a cone diagram of the SDSS wall
galaxies ($100 < r < 260 \mpc$, $M_{r}\lsim -17.5$). The black
points show the void galaxies from the apparent magnitude-limited
sample ($r<17.5$). We only plot galaxies with $|\delta| \lsim
15^{\circ}$. Note that some of the black points appear to be close to
magnitude-limited galaxies. This is, however, just a projection effect (we suppress
the $z$ direction) and all void galaxies have less than three
neighbors within a 3-dimensional radius of $7\mpc$.}
\label{fig:vgals}
\end{centering}
\end{figure}

Figure \ref{fig:vgals} shows a redshift cone diagram of the SDSS
wall galaxies (gray dots) and the corresponding void
galaxies, VGD (black points). We plot only galaxies with $|\delta|
\lsim 15^{\circ}$.  Note that some of the void galaxies appear to be
close to wall galaxies. This is merely a projection effect.  All the
void galaxies have less than three neighbors within a radius of
$7\mpc$.

\begin{figure}
\begin{centering}
\begin{tabular}{c}
{\epsfxsize=11truecm \epsfysize=11truecm \epsfbox[35 170 550 675 ]{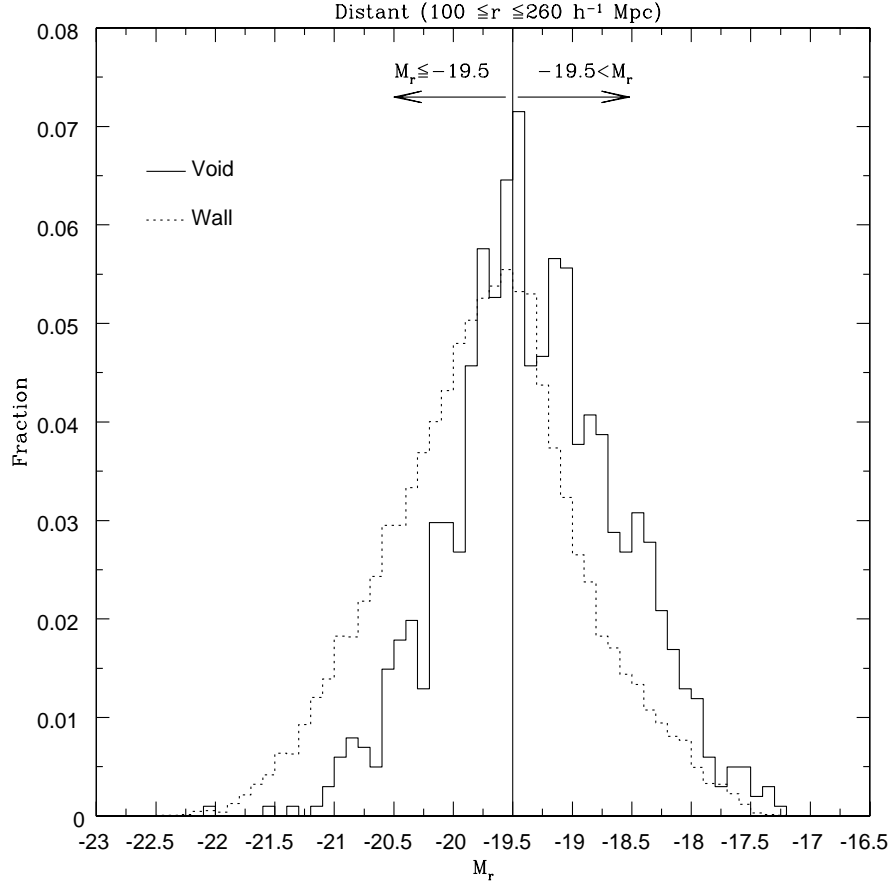}}
\end{tabular}
\caption{Distribution of the absolute magnitudes in the distant wall
(WGD, dotted line) and void (VGD, solid line) galaxy samples. The
parent sample is apparent magnitude-limited at $r\leq17.5$
and redshift limited at $z\leq0.089$. We split both data sets at
$M_{r}=-19.5$ to obtain the void galaxy sub-samples [VGD\_b
(b=bright), VGD\_f (f=faint)] and wall galaxy sub-samples [WGD\_b,
WGD\_f]. The cut at $M_r=-19.5$ divides both the void and wall galaxy
samples into approximately equal halves. This histogram bins galaxies
by $\Delta M=0.1$. The range of absolute magnitudes probed by the void
galaxies that lie within this volume is $-22.0\lsim M_{r}
\lsim-17.2$.}
\label{fig:vg2}
\end{centering}
\end{figure}

After obtaining the WGD and VGD samples, we split each void and
corresponding wall galaxy sample into approximately equal halves by
applying an absolute magnitude cut. In this case, the magnitude cut is
done at $M_{r}=-19.5$, from which we obtained the corresponding
sub-samples, [WGD\_b, VGD\_b] ($M_{r}\leq-19.5$, b=bright) and
[WGD\_f, VGD\_f] ($M_{r}>-19.5$, f=faint). The approximate range of
absolute magnitudes covered by the sub-samples are, $-22\lsim
M_{r}\leq-19.5$, for the bright and $-19.5<M_{r}\leq-17.77$, for the
faint half. Figure \ref{fig:vg2} shows the distribution of absolute
magnitudes for the distant samples. Note the terms {\it bright} and
{\it faint} in this context are used to describe the sub-samples
relative to their parent sample.
\subsection{Nearby Void Galaxies}
\label{sec:NVG}

To find faint void galaxies, which are present in the SDSS sample only
at small comoving distances, we use the UZC+SSRS2 volume-limited
sample to trace the voids because the slice-like SDSS samples are too
thin to detect three-dimensional voids in this nearby volume.

The number of galaxies in the SDSS nearby sample, after applying the
boundary corrections, is 2456.  We measure the distance to the third
nearest UZC+SSRS2 volume-limited galaxy and obtain the values
$(\bar{d}_{sep}^{(3)},\sigma_{sep}^{(3)})=(3.9, 1.9) \mpc$, hence the
choice of $d_{\rm crit}=7\mpc$ is still applicable.  In this case we
find 194 void galaxies. We refer to this void galaxy sample as VGN (N
for nearby) and the respective parent wall galaxy sample (after
removing the respective void galaxies) as WGN.

\begin{figure}
\begin{centering}
\begin{tabular}{c}
{\epsfxsize=11truecm \epsfysize=11truecm \epsfbox[35 170 550 675 ]{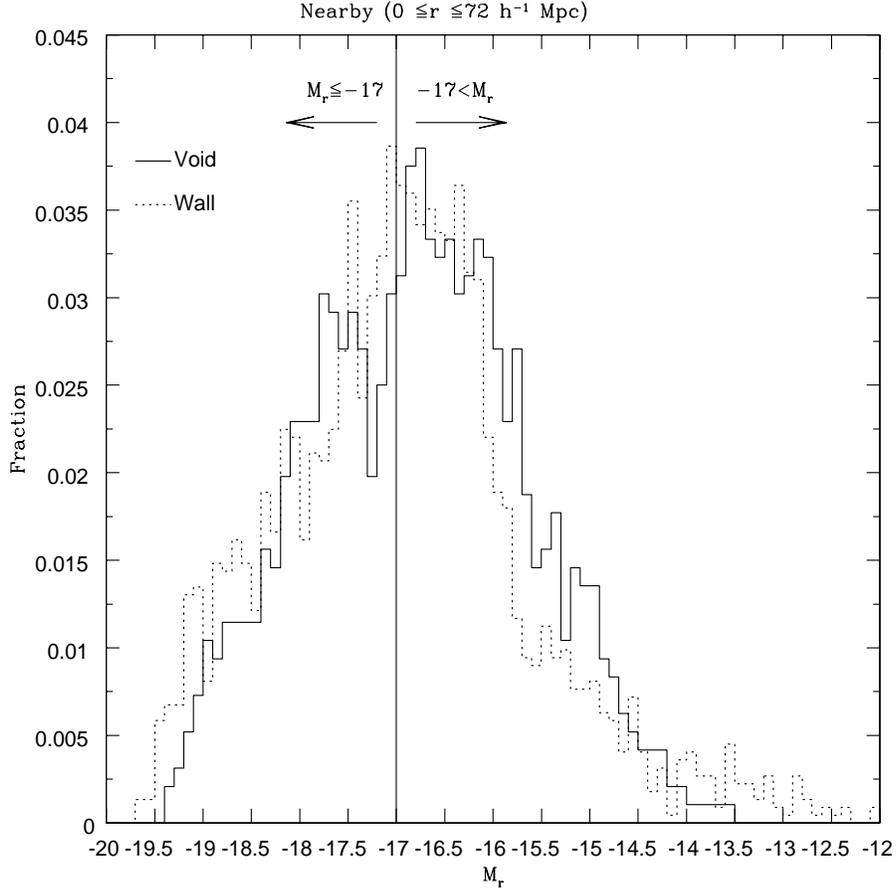}}
\end{tabular}
\caption{Distributions of absolute magnitudes in the nearby wall (WGN,
dotted line) and void (VGN, solid line) galaxy samples. The parent
sample is apparent magnitude-limited at $r\leq17.5$ and
redshift limited at $z\leq0.025$. Both data sets are split at
$M_{r}=-17$ to obtain the void galaxy sub-samples [VGN\_b (b=bright),
VGN\_f (f=faint)] and wall galaxy sub-samples [WGN\_b, WGN\_f]. The
cut at $M_r=-17$ divides both the void and wall samples into
approximately equal halves. This histogram bins galaxies by $\Delta
M=0.1$. The range of absolute magnitudes probed by the void galaxies
that lie within this volume is $-19.7\lsim M_{r} \lsim-13.0$.}
\label{fig:vg1}
\end{centering}
\end{figure}

We again apply an absolute magnitude cut to the VGN and WGN
samples. For the nearby sample, this cut is done at $M_{r}=-17.0$ (see
Figure \ref{fig:vg1}). This cut divides the wall and respective void
galaxy samples into approximately equal halves which we label [WGN\_b,
VGN\_b] ($M_{r}\leq-17.0$, b=bright) and [WGN\_f, VGN\_f]
($M_{r}>-17.0$, f=faint). The range of absolute magnitudes included in
each sub-sample is $-19.7\lsim M_{r}\leq-17.0$ for the bright half
and $-17.0<M_{r}\lsim-13.0$ (see
Figure \ref{fig:vg1}), for the faint half. In this case the
percent of void galaxies found is $8.6\%$.

\section{Photometric Properties}
\label{sec:props}

To examine whether void and wall galaxies have different photometric
properties, we compare their colors ($u-r$ and $g-r$), concentration
indices, and Sersic indices. We compare the properties of wall and
void galaxies in both the distant and nearby samples. We also
subdivide each sample by absolute magnitude and compare their
properties further. The samples compared are therefore, (1) Distant;
bright ($M_{r}\leq-19.5$) [WGD\_b, VGD\_b], faint ($M_{r}>-19.5$)
[WGD\_f, VGD\_f], and full (undivided) void vs. wall samples in each
case respectively and (2) Nearby; bright ($M_{r}\leq-17.0$) [WGN\_b,
VGN\_b], faint ($M_{r}>-17.0$) [WGD\_f, VGD\_f], and full (undivided)
void vs. wall samples in each case respectively.

We compute the means of the distributions and also the error on the
mean to see if on average void and wall galaxies have the same colors,
concentration indices and Sersic indices. We also use the {\it
Kolmogorov-Smirnov} (KS) test to see if the void and wall galaxies
could be drawn from the same parent population. 

Tables 1 and 2, summarize the results of these tests for the nearby
and distant samples respectively. We present the results for the whole
sample, as well as the samples split by absolute magnitude. The
results are considered in detail below.

\subsection{Color}

The existence of strong correlations of galaxy type with density
(Postman \& Geller 1984; Dressler 1980), galaxy type with color
(Strateva et al. 2001; Baldry et al. 2003), and density with
luminosity and color (Hogg et al. 2002; Blanton et al. 2002) are well
known; bright red galaxies tend to populate galaxy clusters and tend
to be elliptical, while dim blue galaxies are less clustered and tend
to be more disk like.  This behavior is shown in an analysis of SDSS
galaxy photometry by Blanton et al. (2002; see their Figures 7 and 8),
in which they find that the distribution of $(g-r)$ colors at redshift
0.1 is bimodal.

Of particular interest to us is the location of the void galaxies in
color space. Because these galaxies evolve more slowly and interact
less with neighboring galaxies than their wall galaxy counterparts, we
might expect void galaxies to be dim, blue, of low-mass and have high
star formation rates (Benson et al. 2003).  We consider two color
indices: $u-r$ and $g-r$. The reason for these two colors is that
$g-r$ measures the slope of the spectrum and $u-r$ is sensitive to the
UV flux and the $4000{\rm \AA}$ break. Since the $u$ band magnitudes
can be noisy, by looking at $g-r$ and $u-r$ we are able to verify that
the results are consistent and not affected by low signal-to-noise
ratio.

\begin{figure}
\begin{centering}
\begin{tabular}{c}
{\epsfxsize=11truecm \epsfysize=11truecm \epsfbox[35 170 550 675 ]{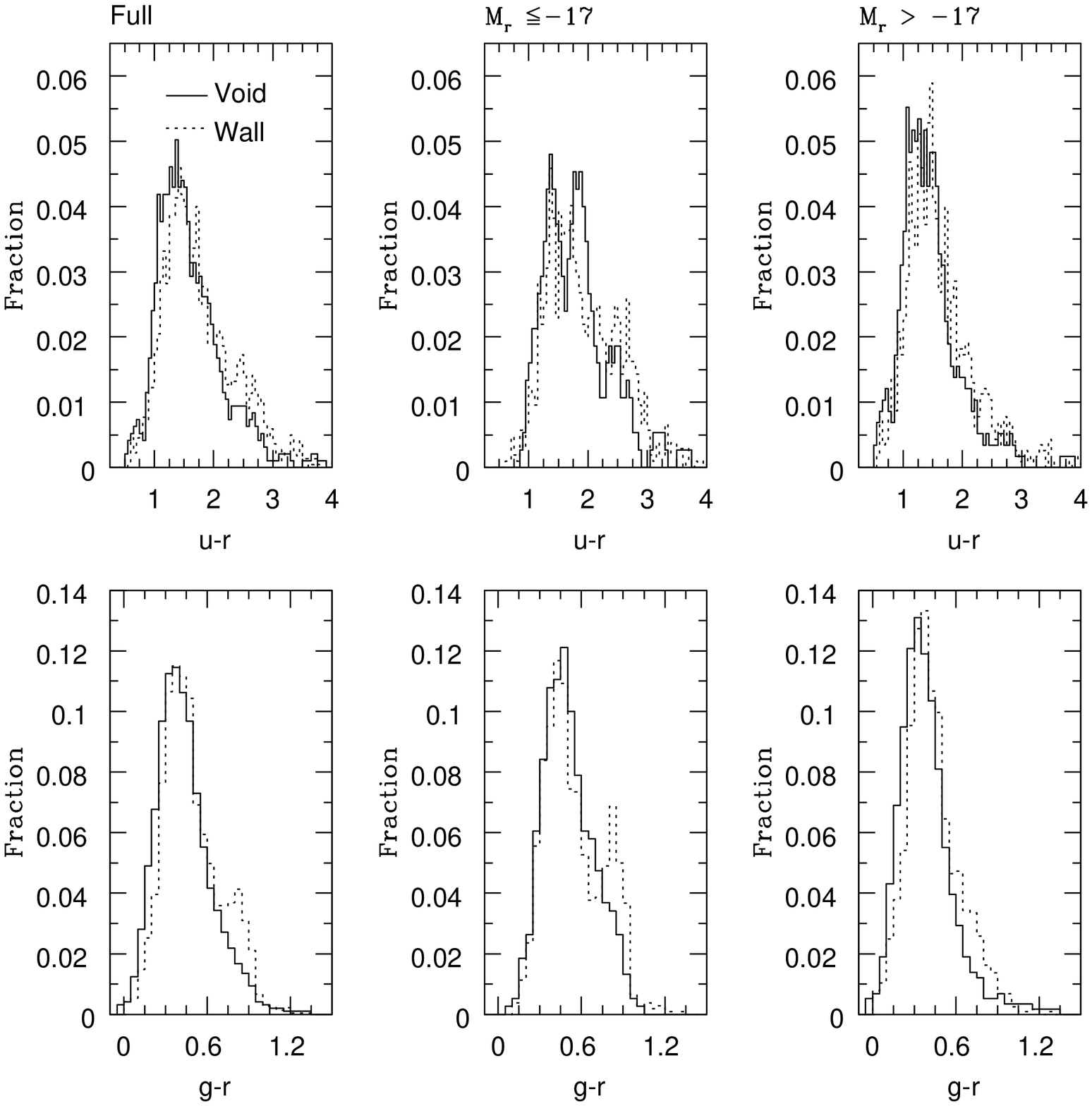}}
\end{tabular}
\caption{Color distributions of nearby void galaxies (solid lines)
compared to the nearby wall galaxies (dotted lines) in two optical
colors, $u-r$ (top row) and $g-r$ (bottom row). The first, second and
third columns are the undivided (full), bright ($M_{r}\leq -17$) and
faint ($-17<M_{r}$) samples respectively. The fraction of galaxies per
0.05 bin of color is shown on the {\rm Y}-axis. In all cases, the
solid curves are shifted to the left i.e., on average, the void
galaxies are bluer than wall galaxies (see Table 1). A KS test reveals
that in the case of the faint sub-sample and full sample it is very
unlikely ($\bar{P}\lsim 0.007$) that the full void and wall galaxy
samples are drawn from the same respective parent populations. In the
bright sub-sample, we see an excess of luminous
red galaxies at $g-r\sim0.9$ that is not present in the void galaxy
histogram.}
\label{fig:color1}
\end{centering}
\end{figure}

\begin{figure}
\begin{centering}
\begin{tabular}{c}
{\epsfxsize=11truecm \epsfysize=11truecm \epsfbox[35 170 550 675 ]{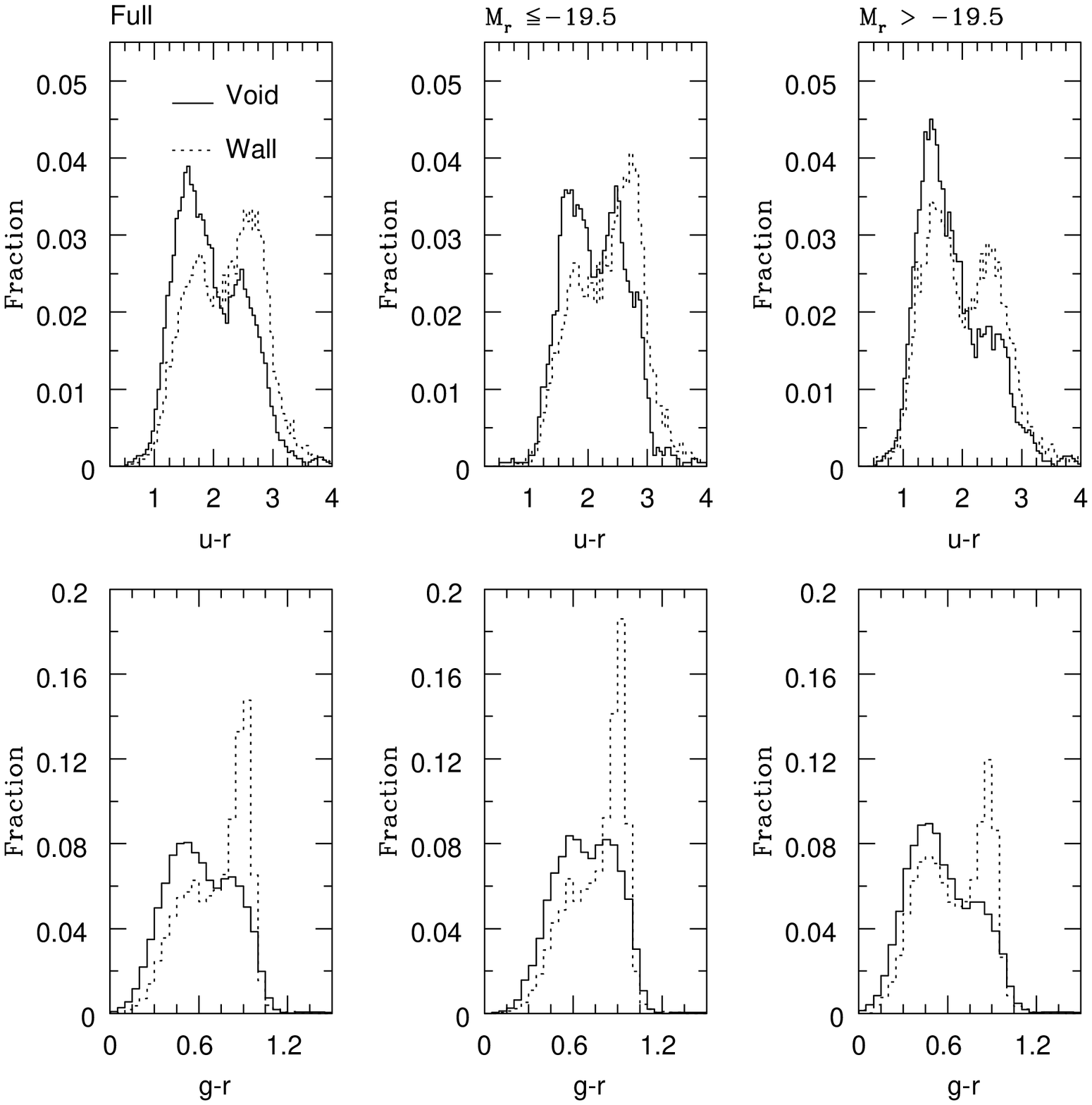}}
\end{tabular}
\caption{Color distributions of distant void galaxies (solid lines)
compared to the distant wall galaxies (dotted lines) in two optical
colors, $u-r$ (top row) and $g-r$ (bottom row). The first, second and
third columns are the undivided (full), bright ($M_{r}\leq-19.5$)
and faint ($-19.5<M_{r}$) samples respectively. The fraction of
galaxies per 0.05 bin of color is shown on the {\rm Y}-axis. In all
cases, the solid curves are shifted to the left i.e., on average, the
void galaxies are bluer than wall galaxies (see Table 2.). A KS test
reveals that it is very unlikely ($P < 10^{-4}$) that the 
void galaxy and wall galaxy samples are drawn from the same parent
populations. We clearly see an excess of luminous,
red galaxies in the wall galaxy histograms as a peak at $g-r\sim1.0$.}
\label{fig:color2}
\end{centering}
\end{figure}

In Tables 1 and 2, we compare the photometric properties of the void
galaxy samples to their respective wall galaxy samples. In Figures
\ref{fig:color1} and \ref{fig:color2} we present normalized histograms
of the color distributions. Solid lines correspond to the void galaxy
samples and the dotted lines represent the wall galaxies. In all cases
(nearby, distant and the bright and faint sub-samples) we find that
the void galaxy samples are on average bluer than the corresponding
wall galaxy samples in both colors. If we look at the full samples, we
find that the mean values of the two samples are significantly
different. The nearby void galaxies have mean $u-r$ and $g-r$ colors
that are at least $3 \sigma_{\mu}$ bluer than the wall galaxy
samples. For the distant void galaxies, the differences in the means
are about four times greater than for the nearby case.

When we split the nearby sample into the bright and faint samples, we
see that it is at the faint end where there is the greatest difference
between void and wall galaxies. The significance of the KS test is
reduced because of the smaller number of galaxies in each sample. The
nearby bright and faint void galaxies are at least $2 \sigma_{\mu}$
bluer than the wall galaxies. The differences between the nearby void
and wall galaxies are not as pronounced as in the distant samples
because we are shot noise limited by how many clusters there are in
the small nearby volume.  In the distant sample it is very unlikely
that the wall and void galaxies in both the bright and faint
sub-samples are drawn from the same parent population ($P<10^{-4}$).

We assess the statistical significance of differences in the color
distributions using a KS test (the values of $P$, the probability that
the two samples are drawn from the same parent population, are given
in the last column of Tables 1 and 2). The probability that the void
and wall galaxies are drawn from the same parent population is low:
$P<0.002$ in the nearby case and $P<10^{-4}$ in the distant case. In
the distant sample it is very unlikely that the wall and void galaxies
in both the bright and faint sub-samples are drawn from the same
parent population ($P<10^{-4}$).

\subsection{Concentration Index}

To compare morphological properties of void and wall galaxies, we
examine the distribution of concentration indices measured by the SDSS
photometric pipeline (Lupton et al. 2001; Stoughton et al. 2002; Pier
et al. 2002; Lupton et al. 2002).  The concentration index (CI) is
defined by the ratio $\rm {CI}\equiv r90/r50$, where $r90$ and $r50$
correspond to the radii at which the integrated fluxes are equal to
$90\%$ and $50\%$ of the Petrosian flux, respectively. A large value
of CI corresponds to a relatively diffuse galaxy and a small value of
CI to a highly concentrated galaxy. The concentration index has been
shown to correlate well with galaxy type (Strateva et al. 2001;
Shimasaku et al. 2001). Spiral galaxies are usually found to have
small concentration indices ( $\lsim 2.5$) whereas ellipticals have
larger concentration indices ($\gsim 2.5$). This bimodal behavior of
the concentration index can be clearly seen in Strateva et al. (2001;
see Figure 8).

\begin{figure}
\begin{centering}
\begin{tabular}{c}
{\epsfxsize=11truecm \epsfysize=11truecm \epsfbox[35 170 550 675 ]{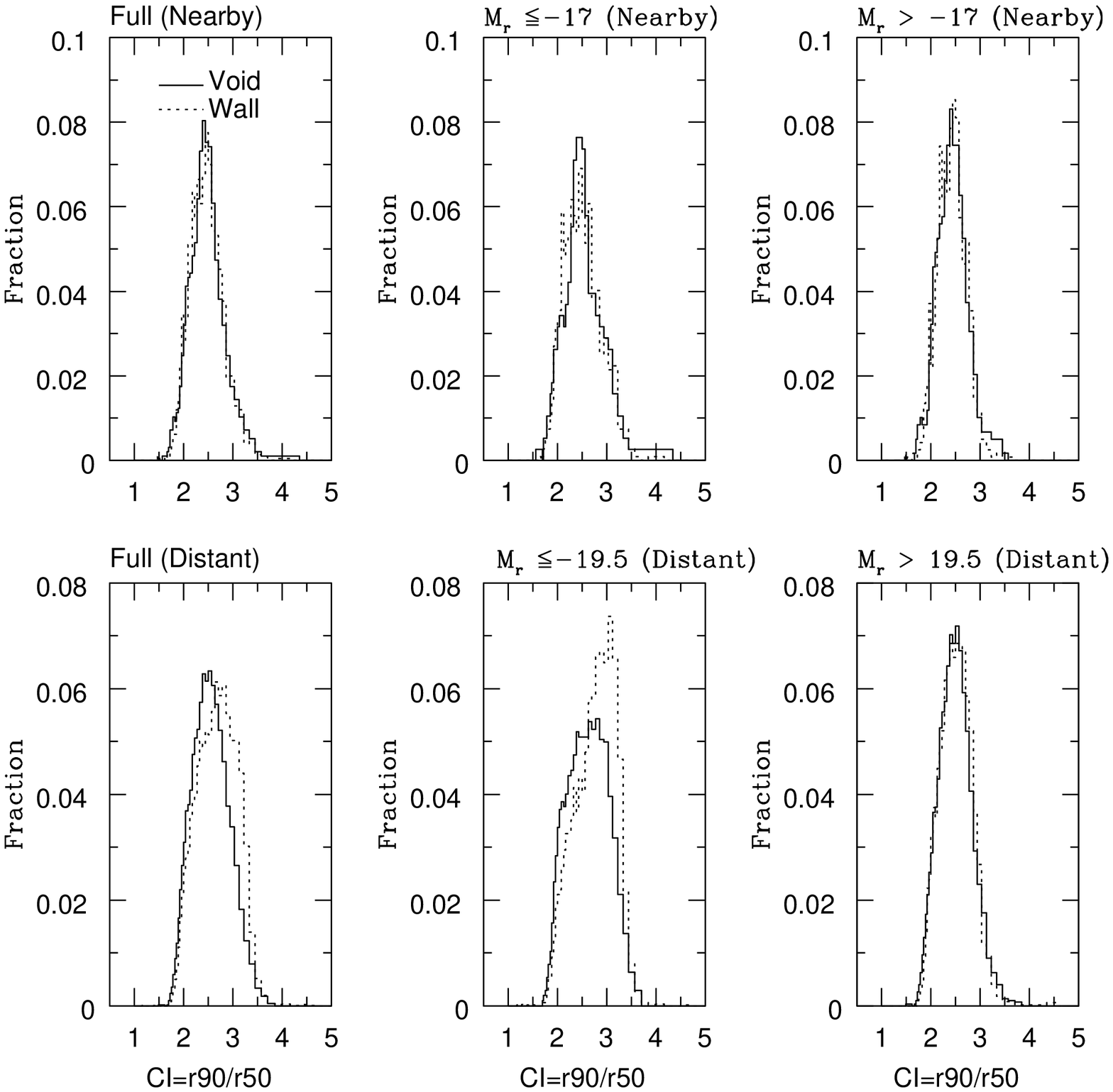}}\\
\end{tabular}
\caption{Concentration Index Distribution. We show the normalized
fraction of void (solid lines) and wall galaxies (dotted lines) as a
function of $r90/r50$. The top row shows the results for the nearby
galaxies, the bottom row shows the results for the distant galaxies.
The first, second and third columns are again the full, bright and
faint samples. The fraction of galaxies per 0.1 bin of concentration
index is shown on the {\rm Y}-axis. The shift in the distribution
for the distant, bright ($M_{r}\leq-19.5$) wall galaxies around
$CI\sim2.5$ corresponds to the excess of bright red galaxies in the
wall samples. The KS
statistic reveals that the distant void galaxy (bright and full)
and respective wall galaxy samples are very different from one
another, with a probability of $<0.01\%$ that they are drawn
from the same parent population. In the case of the nearby galaxies,
the two distributions are indistinguishable.}
\label{fig:cin}
\end{centering}
\end{figure}

Figure \ref{fig:cin} shows histograms of CI for void and wall
galaxies for both the nearby and distant samples along with the
respective bright and faint sub-samples. Tables 1 and 2 show the
mean, error on the mean, and the KS statistic found when comparing
the wall and void galaxies.

In the nearby samples, the void and wall galaxies are not
distinguished by this morphological parameter. In Table 1, we find
that the mean values of CI are very similar. The probability
that the distributions of concentration indices of void and wall
sub-samples are drawn from the same parent population approaches unity
and $0.5$ for the faint and bright sub-samples respectively. The top row
of plots in Figure \ref{fig:cin}, shows there is indeed little
difference between the distributions of CI for the void and wall
galaxies in these samples.

We find that void galaxies have on average significantly smaller
concentration indices in the bright half of the distant samples. There
are more wall than void galaxies at large values of CI
($\rm{CI}\gsim2.5$). The means differ by more than $7 \sigma_{\mu}$.
In Figure \ref{fig:cin}, in the bottom row, all three dotted curves
show this behavior. In Table 2, it is clear that in the full sample
and in the bright sample, the wall and void galaxies have
significantly different CI distributions. A KS analysis of the bright
sub-sample reveals that there is a probability of less than $10^{-4}$
that the void and wall galaxies are drawn from the same parent
population. In the case of the distant faint void and wall galaxy
samples the results are consistent.

\subsection{Sersic Index}

As another measure of morphology of void and wall galaxies we examine
the Sersic index (Sersic 1968), found by fitting the functional form
$I(r)=I_{o}exp(-r^{1/n})$, where $n$ is the Sersic index itself, to
each galaxy surface brightness profile (SBP).  With this form, $n=1$
corresponds to a purely exponential profile, while $n=4$ is a de
Vaucouleurs profile. 
We use the Sersic indices as measured by Blanton
et al. (2002) for the SDSS galaxies.

\begin{figure}
\begin{centering}
\begin{tabular}{c}
{\epsfxsize=11truecm \epsfysize=11truecm \epsfbox[35 170 550 675 ]{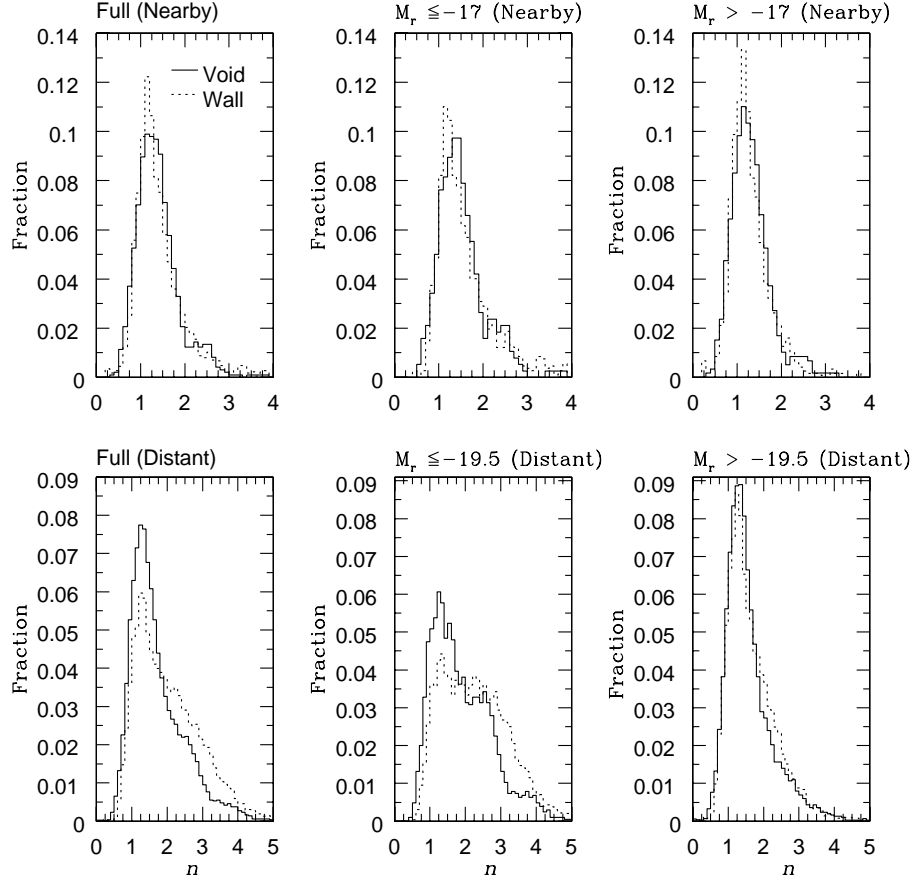}}\\
\end{tabular}
\caption{Sersic Index Distribution. We plot the normalized fraction of
void (solid lines) and wall galaxies (dotted lines) as a function of
Sersic index. The top row shows the results for the nearby galaxies,
the bottom row shows the results for the distant galaxies. The first,
second and third columns are again the full, bright and faint
samples. The fraction of galaxies per 0.1 bin of Sersic index is shown
on the {\rm Y}-axis.  We cannot distinguish the nearby void galaxies
from the respective nearby wall galaxy sample based on the Sersic
index.  However, for the distant samples, the KS test assigns a
probability of $P<10^{-4}$ that the void galaxies are drawn
from the same parent population as the respective wall galaxies. }
\label{fig:nser}
\end{centering}
\end{figure}

In Figure \ref{fig:nser}, we plot histograms of Sersic indices
measured for all the samples. Statistics of these distributions and
the results of comparison of void and wall sub-samples are listed in
Tables 1 and 2.  In the nearby survey volume, we find
$\bar{n}\lsim1.5$ for all void and wall galaxy sub-samples and there
are no statistically significant differences between the
distributions. The top panels of Figure \ref{fig:nser}, show
histograms of the Sersic index; the distributions of the void (solid
lines) and wall (dotted line) galaxies appear very similar.

We find significant differences between void and wall galaxies in the
distant samples.  The lower panels in Figure \ref{fig:nser}, show the
distribution of Sersic indices for the void and wall galaxies.  For
the more distant void galaxies, we find in Table 2, that $\bar{n}_{\rm
Distant}=1.7$, which is higher than what was found for the nearby void
galaxies ($\bar{n}_{\rm Nearby}=1.4$).  A KS test reveals that the
void galaxies are distinct from the wall galaxies with a probability
of $P<10^{-4}$ in the fainter ($M_r>-19.5$), brighter ($M_r<-20.3$)
and full samples that the void and wall galaxies are drawn from the
same parent population. The means of the
Sersic indices of the void and wall galaxies differ by at least $3
\sigma_{\mu}$.
\section{Discussion}
\label{sec:discus}

The above analysis clearly shows that there is a difference in the
photometric properties of void and wall galaxies. Void galaxies are
fainter and bluer than wall galaxies in all cases. Previous
observational studies (e.g., Vennik et al. 1996; Pustilnik et
al. 2002; Popescu et al. 1997) suggested that isolated galaxies in
voids can be distinguished from non-void galaxies based on their color
with a large enough sample. Here we provide such a sample and even
extend the analysis to compare sub-samples of void and non-void
galaxies of similar luminosity and SBP.

Nearby, the question might be raised as to whether it is the faintest
void galaxies that are particularly blue and that these galaxies
dominate the statistics. To test for this, the nearby sample is cut at
-17.0, reducing the range of absolute magnitude in each bin and again
the void galaxies are bluer than the wall galaxies. A further test was
made where the galaxies were divided into bins of $\delta M = 1$ mag
and still the void galaxies are bluer in every bin, thus the
differences in color are not dominated by the tail of the
distribution. Void galaxies are genuinely bluer than wall galaxies of
the same luminosity.

In the distant sample the differences in color are only partly
explained by the paucity of luminous red galaxies in voids. The
average galaxy in the distant sample has an absolute magnitude of
around -19.5 which is more than a magnitude fainter than an $M^*$
galaxy in the $r$-band ($M^*=-20.80$ Blanton et al. 2001). Galaxies
that are thought of as bright red cluster ellipticals are typically
brighter than $M^*$. In the full sample, the faint sample and the
bright sample (and in the $\delta M = 1$ mag test) void galaxies are
still bluer than wall galaxies. 


In Section \ref{sec:nn} we noted a small excess of void galaxies near
the inner boundary of the volume that encloses the distant samples. We
predicted that this might affect the purity of our void galaxy samples
and thereby lower the apparent statistical significance of differences
between the void and wall galaxy populations.  To test for this
effect, we redo selected analyses, to compare the photometric
properties of void and wall galaxies in the range of comoving
coordinate distance from $r=160\mpc$ to $r=260\mpc$, far from the
region where the excess is observed near the $r=100\mpc$ inner
boundary.  We find that the differences between the photometric
properties of void and wall galaxies are indeed larger for galaxies in
this more restricted redshift range. For example, $u-r$, $u-g$, and
$g-r$ the differences rise to $> 7\sigma_{\mu}$. The sense of these
differences is the same as for the larger sample; void galaxies are
bluer and of later type than wall galaxies.  We bother to include the
more nearby, perhaps slightly diluted, void galaxy sample in our full
analysis because it allows us to probe a larger range of absolute
magnitude.  In fact, we find consistency of results in the nearby and
distant samples over the range of absolute magnitude where these
samples overlap. The statistical significance is comparable perhaps
because the nearby samples are relatively smaller, albeit purer. We
expect that the statistical significance of these comparisons will
rise in future, more complete samples from the SDSS.

One might ask if the observed differences in color are simply the
result of the well-known morphology density relation, extrapolated
down to lower densities: blue spiral galaxies are found in low density
environments, while red ellipticals are found in clusters.  This
explanation seem unlikely in the nearby samples, where the surface
brightness profiles of void and wall galaxies are quite similar. Thus,
in the nearby samples, the difference in color is not clearly linked
to morphology.  In the distant samples, however, we see a
morphological difference between the void and wall sample; there are
more elliptical type galaxies in the wall sample.  To test if the
difference in color is caused simply by the paucity of ellipticals in
voids, we divide the distant sample by Sersic index. Blanton et al.
(2002) use $n<1.5$ to represent exponential disks and $n>3$ for the de
Vaucouleurs profiles whereas Vennik et al. (1996) use $n\leq1.5$ for
exponential law fits and $n\geq1.6$ for early type galaxy profile
fitting.  We examine the color distributions of void and wall galaxies
with Sersic index less than 1.8 and greater than 1.8 to approximately
split the sample into spirals and ellipticals. We find that the void
galaxies with both $n<1.8$ and $n>1.8$ are bluer than the wall
galaxies. In $u-r$ and $g-r$ and for both $n<1.8$ and $n>1.8$ the void
galaxies are at least 3$\sigma_{\mu}$ bluer than the wall galaxies,
and for the $n<1.8$, $g-r$ case the difference rises to
7$\sigma_{\mu}$. Again, the samples are divided into bright and faint
sub-samples as well as by Sersic index. The void galaxies are always
bluer than the wall galaxies, although the significance of the KS test
is reduced because of the smaller number of galaxies.

Thus, void galaxies are bluer than wall galaxies even when
compared at similar SBP and luminosities. They are also fainter and
have surface brightness profiles that more closely resemble spirals
than ellipticals. These findings are consistent with predictions of
void galaxy properties from a combination of Semi-Analytic Modeling
and N-body simulations of structure formation in Cold Dark Matter
models (Benson et al. 2003). One of the reasons why void galaxies are
bluer than galaxies in richer environments may be that star formation is
an ongoing process in void galaxies. Galaxies in clusters and groups
have their supply of fresh gas cut
off. Therefore, star formation is suppressed in the wall galaxies.

To illustrate the range of luminosities probed by this study, we
consider which members of the Local Group could have been included in
our samples at the distances probed by the SDSS volume.  Not only the
brightest members of the Local Group (LG), but also Local Group
members like M31 and M33 can be detected in the distant sample, and
fainter ($M_{r}\gsim-16.5$) members of the LG, like the LMC and SMC,
would be included in the nearby sample.

In the nearby sample we can detect faint dwarf ellipticals (dE's),
which is to be expected given that about $80\%$ of the known galaxies
in the LG are dwarfs (Sung et al. 1998; Staveley-Smith, Davies \&
Kinman al. 1992). It is well known that while dE's have exponential
SBP's (Sandage \& Binggeli 1984; Binggeli, Tammann \& Sandage 1987;
Caldwell \& Bothun 1987) they exhibit color gradients that redden
outward (Jerjen et al. 2000; Vader et al. 1988; Bremnes et al. 1998)
and have a uniform color distribution (James 1994; Sung et al. 1998).
Based on their color and other properties, a fraction of the void
galaxies resemble a population of dwarf ellipticals (dE's), which have
a mean $g-r =0.51$ (Kniazev et al. 2003), Typical dE's have Sersic
indices $n\sim1.0$ and $M_{B}\gsim-17$, consistent with our sample of
nearby void galaxies (see Table 1).

\section{Conclusions}
\label{sec:conc}

Using a nearest neighbor analysis, we identify void galaxies in the
SDSS. For the first time we have a sample of $\sim10^{3}$ void
galaxies. These void galaxies span a wide range of absolute
magnitudes, $-13.5>M_r>-22.5$, are found out to distances of
$260\mpc$, and are found in regions of the universe that have density
contrast $\delta \rho / \rho < -0.6$.

In previous studies of properties void galaxies it was suggested
(Vennik et al. 1996; Pustilnik et al. 2002; Popescu et al. 1997) that
void galaxies could be distinguished from non-void galaxies based on
their color, and a hint of them being bluer was observed (Grogin
\& Geller (1999, 2000) from a small sample of void galaxies.
In this paper we present a definitive result
with a sample of $1204$ void galaxies for which the colors,
concentration and Sersic indices are compared against wall galaxies.

Void galaxies are bluer than wall galaxies of the same intrinsic
brightness and redshift distribution down to $M_{r}=-13.5$. We
demonstrate that the difference in colors is not explained by the
morphology-density relation. Nearby, void and wall galaxies have very
similar surface brightness profiles and still the void and wall
galaxies have different colors. In the distant sample the voids and
wall galaxies have different surface brightness profiles. However,
when we divide the populations further by Sersic index, the void
galaxies are still bluer. To test that the differences in color are not
due to the choice of absolute magnitude range, we compare the colors
within narrow bins of absolute magnitude. This reveals that void
galaxies are genuinely blue and that the differences between the
colors are not dominated by extreme objects in the tails of the void
and wall galaxy distributions.

Analysis of surface brightness profiles indicates that void galaxies
are of later type than wall galaxies. Comparison of the Sersic indices
between the distant void and wall galaxy samples including,
sub-samples within a narrow range of luminosities shows that it is
very unlikely ($P<10^{-4}$) that the two samples are drawn from the
same parent population. However, based on the concentration index, it
is only the bright distant void and wall galaxy samples that differ by
significantly.

Our results are in agreement with predictions from semi-analytic
models of structure formation that predict void galaxies should be
bluer, fainter, and have larger specific star formation rates (Benson
et al. 2003).  The differences in color are probably best explained in
terms of star formation. Void galaxies are probably still undergoing
star formation whereas wall galaxies have their supply of gas
strangled as they fall into clusters and groups.

In a separate paper (Paper II; Rojas et al. 2004) we will discuss
analysis of the spectroscopic properties (${\rm H}_\alpha$, [OII]
equivalent widths, and specific star formation rates) of our void
galaxies. Work in progress reveals that the specific star
formation rate of our void galaxies is considerably higher, consistent
with our current findings and predictions.

\section*{Acknowledgments}

Funding for the creation and distribution of the SDSS Archive has
been provided by the Alfred P. Sloan Foundation, the Participating
Institutions, the National Aeronautics and Space Administration,
the National Science Foundation, the U.S. Department of Energy,
the Japanese Monbukagakusho, and the Max Planck Society. The SDSS
Web site is http://www.sdss.org/.

The SDSS is managed by the Astrophysical Research Consortium (ARC)
for the Participating Institutions. The Participating Institutions
are The University of Chicago, Fermilab, the Institute for
Advanced Study, the Japan Participation Group, The Johns Hopkins
University, Los Alamos National Laboratory, the
Max-Planck-Institute for Astronomy (MPIA), the
Max-Planck-Institute for Astrophysics (MPA), New Mexico State
University, University of Pittsburgh, Princeton University, the
United States Naval Observatory, and the University of Washington.

MSV acknowledges support from NSF grant AST-0071201 and a grant from
the John Templeton Foundation. We acknowledge David Goldberg for
useful conversations. We thank the referee for valuable comments.

\clearpage

\begin{table}
\begin{centering}
\begin{tabular}{c}
\bf{NEARBY SAMPLE}
\end{tabular}
\begin{tabular}{cccccccc}

& &  \bf {Full} ($-19.9 \leq M_{r}\leq -14.5$)  &[$N_{V}=194$, $N_{W}=2256$] &\\ \hline
& & Void (VGN)&  Wall (WGN) &  KS ($P$)&  \\ 
Property &  &$\mu\pm\sigma_{\mu}$  & $\mu\pm\sigma_{\mu}$ & Probability  &\\ \hline

$g-r$       & & $0.433\pm0.014$ & $0.490\pm0.004$ & $0.002$ & \\
$u-r$       & & $1.598\pm0.040$ & $1.764\pm0.013$ & $0.001$ & \\
$r90/r50$   & & $2.390\pm0.024$ & $2.390\pm0.007$ & $0.802$ & \\
$n$         & & $1.388\pm0.034$ & $1.456\pm0.004$ & $0.506$ & \\ \hline \\

& &  \bf {Bright} ($M_{r} \le -17.0$)  &[$N_{V}=76$, $N_{W}=1071$] &\\ \hline
& & Void (VGN\_b)&  Wall (WGN\_b) &  KS ($P$)&  \\ 
Property &  &$\mu\pm\sigma_{\mu}$  & $\mu\pm\sigma_{\mu}$ & Probability  &\\ \hline

$g-r$      & & $0.510\pm0.019$ & $0.549\pm0.006$ & $0.207$ & \\
$u-r$      & & $1.810\pm0.063$ & $1.930\pm0.019$ & $0.104$ & \\
$r90/r50$  & & $2.429\pm0.044$ & $2.421\pm0.011$ & $0.424$ & \\
$n$        & & $1.518\pm0.060$ & $1.626\pm0.004$ & $0.605$ & \\ \hline \\

& &  \bf {Faint} ($M_{r} > -17.0$) &[$N_{V}=118$, $N_{W}=1185$] &\\ \hline
& & Void (VGN\_f)&  Wall (WGN\_f) &  KS ($P$)&  \\ 
Property &  &$\mu\pm\sigma_{\mu}$  & $\mu\pm\sigma_{\mu}$ & Probability  &\\ \hline

$g-r$      & & $0.383\pm0.018$ & $0.436\pm0.006$ & $0.003$ & \\
$u-r$      & & $1.459\pm0.047$ & $1.611\pm0.016$ & $0.018$ & \\
$r90/r50$  & & $2.366\pm0.027$ & $2.361\pm0.008$ & $0.918$ & \\
$n$        & & $1.305\pm0.039$ & $ 1.304\pm0.006$ & $0.509$ & \\ \hline

\end{tabular}
\caption{Means, errors on the means and KS test probabilities that the
void and wall galaxies are drawn from the same parent population for
the photometric properties of void and wall galaxies in the nearby
sample ($r<72 \mpc$). The number of galaxies (void and wall) in each
sample and sub-sample are listed next to the magnitude range heading
as [$N_{V}$ (void), $N_{W}$ (wall)]. Small values of $P$ correspond
to a low probability that the two samples are drawn from the
same parent population. The KS test shows that void galaxies appear to
have different colors to wall galaxies. The void galaxies appear bluer
than the respective wall galaxies in all cases, where the average
difference between the means of the colors is about $2 \sigma_{\mu}$.
However, the concentration and Sersic indices are not significantly different.}
\label{tab:prop1}
\end{centering}
\end{table} 

\begin{table}
\begin{centering}
\begin{tabular}{c}
\bf{DISTANT SAMPLE}
\end{tabular}
\begin{tabular}{cccccccc}
& &  \bf {Full} ($-22.5 \le M_{r} \le -17.77$)   & [$N_{V}=1010$, $N_{W}=12732$] &\\ \hline
& & Void (VGD)&  Wall (WGD) &  KS ($P$) &  \\ 
Property &  &$\mu\pm\sigma_{\mu}$  & $\mu\pm\sigma_{\mu}$ & Probability  &\\ \hline

$g-r$       & & $0.615\pm0.007$ & $0.719\pm0.002$ & $<10^{-4}$ & \\
$u-r$       & & $1.958\pm0.018$ & $2.219\pm0.005$ & $<10^{-4}$ & \\
$r90/r50$   & & $2.449\pm0.011$ & $2.571\pm0.004$ & $<10^{-4}$ & \\
$n$         & & $1.718\pm0.024$ & $2.051\pm0.002$ & $<10^{-4}$ & \\ \hline \\

& &  \bf {Bright} ($M_{r} \le -19.5$)  &[$N_{V}=409$, $N_{W}=7831$] &\\ \hline
& & Void (VGD\_b)&  Wall (WGD\_b) &  KS ($P$) &  \\ 
Property &  &$\mu\pm\sigma_{\mu}$  & $\mu\pm\sigma_{\mu}$ & Probability  &\\ \hline

$g-r$       & & $0.686\pm0.009$ & $0.765\pm0.002$ &  $<10^{-4}$ & \\
$u-r$       & & $2.126\pm0.026$ & $2.343\pm0.006$ &  $<10^{-4}$ & \\
$r90/r50$   & & $2.505\pm0.019$ & $2.656\pm0.004$ &  $<10^{-4}$ & \\
$n$         & & $1.908\pm0.042$ & $2.285\pm0.003$ &  $<10^{-4}$ & \\ \hline \\

& &  \bf {Faint} ($M_{r}>-19.5$)  & [$N_{V}=601$, $N_{W}=4901$]&\\ \hline
& & Void (VGD\_f)&  Wall (WGD\_f) &  KS ($P$) &  \\ 
Property &  &$\mu\pm\sigma_{\mu}$  & $\mu\pm\sigma_{\mu}$ & Probability  &\\ \hline

$g-r$       & & $0.567\pm0.009$ & $0.645\pm0.003$ & $<10^{-4}$ & \\
$u-r$       & & $1.844\pm0.024$ & $2.020\pm0.009$ & $<10^{-4}$ & \\
$r90/r50$   & & $2.411\pm0.013$ & $2.435\pm0.005$ & $0.211$   & \\
$n$         & & $1.589\pm0.028$ & $1.677\pm0.003$ & $<10^{-4}$ & \\ \hline

\end{tabular}
\caption{Means, errors on the means and KS test probabilities that the
void and wall galaxies are drawn from the same parent population for
the photometric properties of void and wall galaxies in the distant
sample ($100\leq r \leq260 \mpc$). The number of galaxies (void and
wall) in each sample and sub-sample are listed next to the magnitude
range heading as [$N_{V}$ (void), $N_{W}$ (wall)]. Small values of
$P$ correspond to a low probability that the two samples are drawn
from the same parent population.  In this case, the KS test shows that
the void and wall galaxies are drawn from different populations based
on both color and morphology (concentration index and surface
brightness profile). The differences between the means of the
different parameters measured are on average $> 5 \sigma_{\mu}$,
except for the concentration index in the faint sub-sample, where the
difference is $\sim 2\sigma_{\mu}$. Void galaxies are on average bluer
and more disklike than wall galaxies.}
\label{tab:prop2}
\end{centering}
\end{table}

\clearpage

\end{document}